\documentclass[twocolumn,showpacs,preprintnumbers,amsmath,amssymb]{revtex4}

\usepackage{graphicx}
\usepackage{dcolumn}
\usepackage{bm}
\usepackage{CJK}
\usepackage{color}
\begin{document}
\title{Thermal Rectification in the Nonequilibrium Quantum-Dot-System}
\author{Tian Chen}
 \affiliation{State Key Laboratory of Low
Dimensional Quantum Physics, Department of Physics, Tsinghua University, Beijing 100084,
People¡¯s Republic of China}
\author{Xiang-Bin Wang}
 \email{xbwang@mail.tsinghua.edu.cn}
 \affiliation{State Key Laboratory of Low
Dimensional Quantum Physics, Department of Physics, Tsinghua University, Beijing 100084,
People¡¯s Republic of China}
\affiliation{ Jinan Institute of Quantum Technology, Shandong
Academy of Information and Communication Technology, Jinan 250101,
People¡¯s Republic of China}

\begin{abstract}
Quantum thermal transport in two-quantum-dot system with Dzyaloshinskii-Moriya interaction (DM interaction) has been studied. The sign of thermal rectification can be controlled through changing the energy splitting or the DM interaction strength. The anisotropic term in the system can also affect the sign of rectification. Compared with other proposals [L Zhang et al, Phys. Rev. B 80, 172301 (2009)], our model can offer larger rectification efficiency and show the potential application in designing the polarity-controllable thermal diode with a small size system (N=2). Moreover, quantum correlations of two-quantum-dots are investigated. We find that almost perfect quantum correlations can be obtained in the large temperature bias region, and quantum entanglement is more sensitive to the change of the DM interaction strength than quantum discord.
\end{abstract}



\maketitle
\section{Introduction}
Thermal diode is a very important thermal devices\cite{A. Nitzan, B. Li, N. B. Li}. It has extensive application in phonon information\cite{L. Wang}. The heat current flow can be induced in a unidirectional way, and can be regarded as the basic part in controlling the large-scale heat flow. So far, many proposals have been raised to design a thermal diode by different models, such as Frenkel-Kontorova lattice and spin chains\cite{B. Li, Y. Yan, L. Zhang}. In all these designs, the different lattice sizes will influence the heat current flow significantly between the two reservoirs. In the spin chain model, the polarity of the thermal diode can be controlled by modulating the spin energy gap at different sites or changing the anisotropic interaction strength between the adjacent spins. The efficiency of the thermal diode model can be measured by thermal rectification ($\mathcal{R}$). In the six-spin-chain system, the thermal rectification $\mathcal{R}$ can reach the value from -0.5 to 0.3 by adjusting the system variables, and the different signs of $\mathcal{R}$ means the change of the thermal diode model polarity. However, as studied in Ref. \cite{Y. Yan}, it is not likely to control the thermal diode model polarity in the small system with the spin number $N\leq4$.

Recently, the problems of the interaction between the quantum-dot system and the thermal environment have attracted a lot of attention\cite{R. Hanson, F. Kheirandisha, H. Mohammadia, L. A. Wu_PRA03}, for the quantum-dot system display its advantage in its scalability and long coherence time\cite{R. Hanson}. Due to the quantum tunnel effect in the quantum-dot system, the spin-spin interaction and the spin-orbit interaction will play an important role\cite{F. Kheirandisha, S. J. Akhtarshenas}. Moreover, when the material possesses the reversal asymmetry, the coexistence of the superexchange interaction and the spin-orbit interaction will induce a new interaction so called Dzyaloshinskii-Moriya (DM) interaction\cite{T. Mo, T. Moriya, S. Chutia, A. Pfund, S. Herzog}. It has been verified\cite{T. Mo, T. Moriya} that DM interaction can result in the change of the spin alignment in the material, and many problems of the quantum-dot system containing the DM interaction have been investigated already\cite{F. Kheirandisha, H. Mohammadia, S. J. Akhtarshenas}.

In this paper, we study extensively the heat transport problem in the two-quantum-dot system containing the DM interaction. We calculate the heat current in the steady state limit and analyze the heat current contributions of the different system states. We propose a polarity-controllable thermal diode in this two-quantum-dot system. Thermal rectification of this diode model is studied. We find that, in this two-quantum-dot system, thermal rectification $\mathcal{R}$ can reach the value from -0.3 to 0.35, through changing the DM interaction strength. Compared with the six-spin-chain proposal, thermal rectification $\mathcal{R}$ in our proposal is larger than several cases in the spin-chain proposal. Based on this two-quantum-dot system, we can realize the reversal of thermal rectification in small system case. Besides, the quantum correlations are calculated for the steady state. We find that, compared with the quantum discord, the quantum entanglement is more sensitive to the DM interaction. Perfect quantum entanglement and quantum discord can appear in the large temperature bias region, not only in the two zero reservoirs\cite{F. Kheirandisha}.

\section{Thermal rectification}
\subsection{Model}
\begin{figure}[tbp]
\begin{center}
\includegraphics[width=7cm]{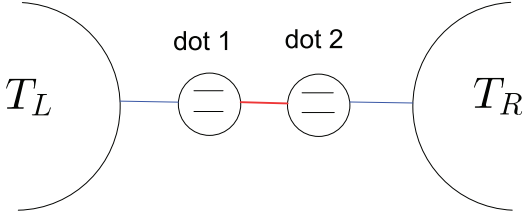}
\end{center}
\caption{\label{fig1} (color online): A schematic representation of the two-quantum-dot system containing DM interaction. The two quantum dots interact with the thermal reservoirs at different temperatures $T_{L}$ and $T_{R}$, respectively.}
\end{figure}

Consider a two-quantum-dot system and two thermal reservoirs as shown in Fig. \ref{fig1}. The left quantum dot interacts with the left reservoir only, and the right quantum dot interacts with the right reservoir only. The total Hamiltonian is\cite{F. Kheirandisha, H. Mohammadia}, $\hat{H}=\hat{H}_{s}+\hat{H}_{BL}+\hat{H}_{BR}+\hat{H}_{SBL}+\hat{H}_{SBR}$. Here, $H_{s}$ is the two-quantum-dot system Hamiltonian, $H_{B\nu}$ is the reservoir Hamiltonian, and $H_{SB\nu}$ is the interaction between the reservoir $\nu$ and the system. We consider the anisotropic interaction and the DM interaction between the two-quantum-dot system, and neglect the second order spin-orbit interaction. The system Hamiltonian is\cite{F. Kheirandisha},
\begin{equation}
\begin{split}
H_{s}=&J\chi(\sigma_{L}^{+}\sigma_{R}^{+}+\sigma_{L}^{-}\sigma_{R}^{-})+J(1+iD)\sigma_{L}^{+}\sigma_{R}^{-}\\
&+J(1-iD)\sigma_{L}^{-}\sigma_{R}^{+}+\frac{B+b}{2}\sigma_{L}^{z}+
\frac{B-b}{2}\sigma_{R}^{z}\\
\end{split}
\end{equation}
where, the parameter $\chi$ describes the anisotropy term in the system spin xy plane, the parameter D stands for the DM interaction between the two-quantum-dot system, and the energy gaps of the left spin and the right spin are $\frac{B+b}{2}$, $\frac{B-b}{2}$ respectively.

In our case, the reservoir consists of a collection of non-interacting oscillators. The parameter $\omega_{n}$ labels different oscillator frequency. The system-reservoir interaction strength $g_{n}^{\nu}$ denotes the interaction between the spin $\nu$ and the oscillator of frequency $\omega_{n}$ in reservoir $\nu$. The reservoir Hamiltonian $H_{B\nu}$ and the system-reservoir interaction Hamiltonian $H_{SB_{\nu}}$ can be written as\cite{F. Kheirandisha},
\begin{equation}
\begin{split}
H_{B\nu}&=\sum_{n}\omega_{n}b_{n\nu}^{+}b_{n\nu}\quad\\ H_{SB\nu}&=S_{\nu}B_{\nu}=(\sigma_{\nu}^{+}+\sigma_{\nu}^{-})(\sum_{n}g_{n}^{(\nu)}b_{n\nu}+g_{n}^{(\nu)*}b_{n\nu}^{+})
\end{split}
\end{equation}

We want to investigate the system properties in the steady state limit. The eigenvalues and the corresponding eigenvectors of the system Hamiltonian are\cite{F. Kheirandisha},
\begin{equation}
\begin{split}
|\varepsilon_{1}\rangle&=N^{+}((\frac{b+\xi}{J(1-iD)})|01\rangle+|10\rangle), E_{1}=+\xi\\
|\varepsilon_{2}\rangle&=N^{-}((\frac{b-\xi}{J(1-iD)})|01\rangle+|10\rangle), E_{2}=-\xi\\
|\varepsilon_{3}\rangle&=M^{+}((\frac{B+\eta}{J\chi})|00\rangle+|11\rangle), E_{3}=+\eta\\
|\varepsilon_{4}\rangle&=M^{-}((\frac{B-\eta}{J\chi})|00\rangle+|11\rangle), E_{4}=-\eta\\
\end{split}
\end{equation}
Here, the state $|0\rangle$ and $|1\rangle$ are the spin of upper level energy and spin of lower level energy, respectively. The coefficients $N^{\pm}$, $M^{\pm}$, $\xi$, and $\eta$ satisfy the relation $N^{\pm}=(1+\frac{(b\pm\xi)}{J^{2}+(JD)^{2}})^{-1/2}$, and $M^{\pm}=(1+(\frac{B\pm\eta}{J\chi})^{2})^{-1/2}$, $\xi=(b^{2}+J^{2}+(JD)^{2})^{1/2}$, $\eta=(B^{2}+(J\chi)^{2})^{1/2}$. We now rewrite the system components in the system-reservoir interaction Hamiltonian $H_{SB\nu}$ in the new basis $\{|\varepsilon_{1}\rangle, |\varepsilon_{2}\rangle, |\varepsilon_{3}\rangle, |\varepsilon_{4}\rangle\}$,
\begin{equation}
\begin{split}
\sigma_{L}^{x}=&S_{3,1}^{L}|\varepsilon_{3}\rangle\langle\varepsilon_{1}|+S_{4,1}^{L}|\varepsilon_{4}\rangle\langle
\varepsilon_{1}|\\&+S_{3,2}^{L}|\varepsilon_{3}\rangle\langle\varepsilon_{2}|+S_{4,2}^{L}|\varepsilon_{4}\rangle\langle\varepsilon_{2}|+H.C.\\
\sigma_{R}^{x}=&S_{3,1}^{R}|\varepsilon_{3}\rangle\langle\varepsilon_{1}|+S_{4,1}^{R}|\varepsilon_{4}\rangle\langle\varepsilon_{1}|\\&+
S_{3,2}^{R}|\varepsilon_{3}\rangle\langle\varepsilon_{2}|+S_{4,2}^{R}|\varepsilon_{4}\rangle\langle\varepsilon_{2}|+H.C.\\
\end{split}
\end{equation}
The parameters $|S_{m,n}^{\nu}|$ is defined as, $|S_{3,1}^{L}|^{2}=|S_{4,2}^{L}|^{2}=\frac{\eta\xi-Bb}{2\eta\xi}+\frac{J^{2}\chi}{2\eta\xi}, |S_{4,1}^{L}|^{2}=|S_{3,2}^{L}|^{2}=\frac{\eta\xi+Bb}{2\eta\xi}-\frac{J^{2}\chi}{2\eta\xi}, |S_{3,1}^{R}|^{2}=|S_{4,2}^{R}|^{2}=\frac{\eta\xi+Bb}{2\eta\xi}+\frac{J^{2}\chi}{2\eta\xi}, \quad |S_{4,1}^{R}|^{2}=|S_{3,2}^{R}|^{2}=\frac{\eta\xi-Bb}{2\eta\xi}-\frac{J^{2}\chi}{2\eta\xi}$.

We apply the master equation method to solve the system dynamics. We obtain the system time evolution function by the second order approximation in the interaction picture\cite{L. A. Wu, L. A},
\begin{equation}
\begin{split}
\frac{d\rho_{m,n}}{dt}=&-i[H_{SBj}(t),\rho(0)]_{m,n}\\&-\int_{0}^{t}d\tau[H_{SBj}(t),[H_{SBj}(\tau),\rho(\tau)]]_{m,n}
\end{split}
\end{equation}
In our calculation, we assume the weak interaction of the system and the reservoirs. The Markov approximation is applied, and the reservoirs keep in its thermal equilibrium throughout the evolution. Our goal is to see the system steady state behavior in the long time limit, we can neglect the coherence term in the system. Under these conditions mentioned above, we can get the Pauli master equation related to the system state population $\rho_{nn}$ (n=1, 2, 3, 4),
\begin{equation}
\dot{\rho_{nn}}^{s}=-\sum_{m}W_{nm}\rho_{nn}^{s}+\sum_{m}W_{mn}\rho_{mm}^{s}\\
\end{equation}
Where $W_{nm}$ is $W_{nm}=k_{n\rightarrow m}^{L}|S_{m,n}^{L}|^{2}+k_{n\rightarrow m}^{R}|S_{m,n}^{R}|^{2}$, and the rates $k_{n\rightarrow m}$ are,
\begin{equation}
\begin{split}
k_{n\rightarrow m}^{\nu}&=\Gamma_{B,\nu}(\omega_{nm})(n_{B}^{\nu}(\omega_{nm})+1),\quad n>m\\
k_{n\rightarrow m}^{\nu}&=\Gamma_{B,\nu}(\omega_{mn})n_{B}^{\nu}(\omega_{mn}),\qquad n<m
\end{split}
\end{equation}
Here, $n_{B}^{\nu}(\omega)$ denotes thermal distribution of the reservoir $\nu$ at temperature $T_{\nu}$, i.e. $n_{B}^{\nu}(\omega_{pq})=\frac{1}{e^{\omega_{pq}/T_{\nu}}-1}$. $\Gamma_{B,\nu}(\omega_{pq})=2\pi\sum_{n}g_{n,\nu}^{2}\delta(\omega_{pq}-\omega_{n})$ is the interaction strength between the reservoir $\nu$ and the system. The system steady state population are,
\begin{equation*}
\begin{split}
\rho_{11}^{s}=\frac{W_{41}W_{31}}{(W_{14}+W_{41})(W_{13}+W_{31})}\\
\rho_{22}^{s}=\frac{W_{13}W_{14}}{(W_{14}+W_{41})(W_{13}+W_{31})}\\
\end{split}
\end{equation*}
\vspace{1cm}
\begin{equation}
\begin{split}
\rho_{33}^{s}=\frac{W_{41}W_{13}}{(W_{14}+W_{41})(W_{13}+W_{31})}\\
\rho_{44}^{s}=\frac{W_{31}W_{14}}{(W_{14}+W_{41})(W_{13}+W_{31})}
\end{split}
\end{equation}
with the basis $\{|\varepsilon_{1}\rangle, |\varepsilon_{2}\rangle, |\varepsilon_{3}\rangle, |\varepsilon_{4}\rangle\}$. Expand the density matrix $\rho^{s}$ in the system basis $\{|00\rangle,|01\rangle,|10\rangle,|11\rangle\}$, the system reduced density matrix $\rho^{s}$ is,

\begin{widetext}
\begin{equation}
\left(
\begin{array}{cccc}
\rho_{33}^{s}\frac{\eta+B}{2\eta}+\rho_{44}^{s}\frac{\eta-B}{2\eta} & 0 & 0 & \rho_{33}^{s}\frac{J\chi}{2\eta}-\rho_{44}^{s}\frac{J\chi}{2\eta}\\
0 & \rho_{11}^{s}\frac{\xi+b}{2\xi}+\rho_{22}^{s}\frac{\xi-b}{2\xi} & \rho_{11}^{s}\frac{J(1+iD)}{2\xi}-\rho_{22}^{s}\frac{J(1+iD)}{2\xi} & 0\\
0 & \rho_{11}^{s}\frac{J(1-iD)}{2\xi}-\rho_{22}^{s}\frac{J(1-iD)}{2\xi} & \rho_{11}^{s}\frac{\xi-b}{2\xi}+\rho_{22}^{s}\frac{\xi+b}{2\xi} & 0\\
\rho_{33}^{s}\frac{J\chi}{2\eta}-\rho_{44}^{s}\frac{J\chi}{2\eta} & 0 & 0 & \rho_{33}^{s}\frac{\eta-B}{2\eta}+\rho_{44}^{s}\frac{\eta+B}{2\eta}
\end{array}
\right)
\end{equation}
\end{widetext}

Given the system steady states obtained above, we can now formulate the expression of the heat current.

\subsection{Results}
\textit{Heat current}. The symmetric expression of the heat current $\hat{J}$ is defined as\cite{L. A. Wu}
\begin{equation}
\begin{split}
J=Tr[\hat{J}\rho]\quad \hat{J}=\frac{i}{2}[H_{SBL},H_{s}]+\frac{i}{2}[H_{s},H_{SBR}]
\end{split}
\end{equation}
in our case, the heat current can be written as
\begin{equation}
\begin{split}
J=&\frac{1}{2}\sum_{m,n}\omega_{m,n}|S_{n,m}^{L}|^{2}P_{n}[k_{n\rightarrow m}^{L}(T_{L})]\\&-\frac{1}{2}\sum_{m,n}\omega_{m,n}|S_{n,m}^{R}|^{2}P_{n}[k_{n\rightarrow m}^{R}(T_{R})]
\end{split}
\end{equation}
and $P_{n}=Tr_{Bj}[\rho_{n,n}(t)]$.

\begin{figure}[!htbp]
\begin{center}
\includegraphics[width=0.50\textwidth]{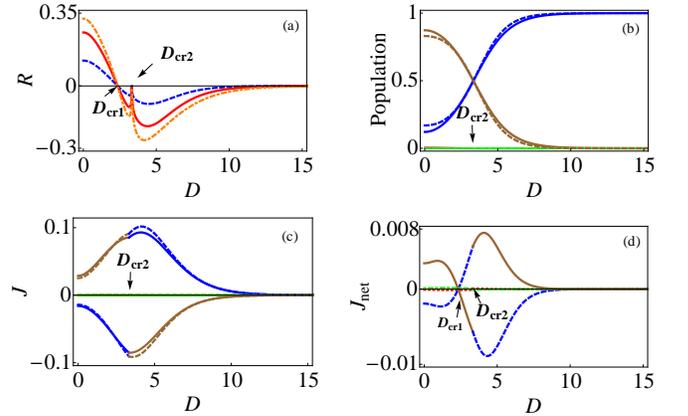}
\end{center}
\caption{\label{fig2} Thermal rectification (a) Thermal rectification with the DM
interaction strength. The sign changes at critical value $D_{cr1}$. Dashed, solid and dot-dashed correspond to $|\Delta| T=0.4$,
$|\Delta| T=1$ and $|\Delta| T=1.6$. The mean temperature of two reservoirs is $TM=1$. (b) The population of four system eigenstates with the change of DM interaction strength. (c) Values of $J_{+,n}$ and $J_{-,n}$ with DM interaction strength. In (b) and (c), solid lines are for the case $\Delta T>0$; (i.e., the temperature of left reservoir is higher than the right one); dashed line are for $\Delta T<0$. The red, blue, green and brown lines correspond to the state $|\varepsilon_{1}\rangle$, $|\varepsilon_{2}\rangle$, $|\varepsilon_{3}\rangle$ and $|\varepsilon_{4}\rangle$ respectively. (d) Values of $J_{net,n}$. The red dotted, blue dashed, green dot-dashed and brown solid line correspond to the state $|\varepsilon_{1}\rangle$, $|\varepsilon_{2}\rangle$, $|\varepsilon_{3}\rangle$, and $|\varepsilon_{4}\rangle$ respectively. $\Delta T=0.4$ and $TM=1$ in (b), (c) and (d). Other parameters are: $\chi=0.3$, B=4, b=2, J=1, $\Gamma_{L}=1$, and $\Gamma_{R}=0.25$.}
\end{figure}

\textit{Thermal rectification control by DM interaction}. Under different temperature bias ($\Delta T=T_{L}-T_{R}$) case, the control of the thermal rectification with system DM interaction strength is shown in the Fig. \ref{fig2}.(a). The mean temperature of two reservoirs is, $TM=1/2(T_{L}+T_{R})$. As defined by Ref.\cite{L. Zhang}, thermal rectification $\mathcal{R}$ satisfies the relation,

\begin{equation}
\mathcal{R}=(J_{+}-J_{-})/\max(J_{+},J_{-})
\end{equation}
where the forward heat current $J_{+}$ is the heat current (from left to right) when the reservoir at higher temperature is connected to the left quantum dot, and the backward current $J_{-}$ is the heat cuurent (from right to left) when the left quantum dot is in contact with the reservoir at lower temperature. The energy gaps of the two quantum dots are not equal. From Fig. \ref{fig2}. (a), the thermal rectification $\mathcal{R}$ will change from the positive value to the negative value after the DM interaction strength reaches a critical value $D_{cr1}$. This means one can obtain rather different heat current through adjusting the DM interaction strength. The different system states population with the change of the DM interaction strength are shown in Fig. \ref{fig2}. (b). When the DM interaction strength is not strong, the system populates mainly at the lowest eigenstate $|\varepsilon_{4}\rangle$. The population of the eigenstate $|\varepsilon_{2}\rangle$ rises with the DM interaction strength. Values of $J_{+,n}$ and $J_{-,n}$ are shown in Fig. \ref{fig2}. (c). To illustrate the contribution of four different system states to the thermal rectification, Define $J_{net}$ as,
\begin{equation}
J_{net,n}=J_{+,n}-J_{-,n}
\end{equation}
where $J_{+,n}$ is the heat current contributed by the system state $|\varepsilon_{n}\rangle$ (from left to right) when the left quantum dot is in contact with the reservoir at higher temperature, and $J_{-,n}$ is the heat current contributed by the system state $|\varepsilon_{n}\rangle$ (from right to left) when the left quantum dot is connected with the reservoir at lower temperature. We display the values of $J_{net}$ in Fig. \ref{fig2}. (d), in the region of the weak DM interaction strength, state $|\varepsilon_{2}\rangle$ provides the negative net heat current, and the net heat current contributions from state $|\varepsilon_{4}\rangle$ is positive. The heat current contribution from state $|\varepsilon_{4}\rangle$ is more notable than that from state $|\varepsilon_{2}\rangle$. The heat current contributed by state $|\varepsilon_{1}\rangle$ and $|\varepsilon_{3}\rangle$ are negligence. So according to the formula of the thermal rectification above, the thermal rectification value is positive when the DM interaction strength is small. After the DM interaction strength reaches the critical value $D_{cr1}$, the contribution of $|\varepsilon_{2}\rangle$ $J_{net,2}$ is positive, and the heat current of $|\varepsilon_{4}\rangle$ $J_{net,4}$ is negative. The major contribution is from $|\varepsilon_{4}\rangle$, and the sign of thermal rectification will change to the negative sign. In our two-quantum-dot system, the lowest system eigenstate changes from state $|\varepsilon_{4}\rangle$ to state $|\varepsilon_{2}\rangle$ when the DM interaction strength reaches the critical value $D_{cr2}$. We find that the net heat current contribution of state $|\varepsilon_{2}\rangle$ is negative, and state $|\varepsilon_{4}\rangle$ provides a positive contribution. When DM interaction exceeds critical value $D_{cr2}$, $J_{net}$ is mainly contributed by state $|\varepsilon_{2}\rangle$. From the analysis above, we find that the lowest eigenstate of the two-quantum-dot system makes the major contribution on the thermal rectification. Through modulating the DM interaction strength, we can change the sign of thermal rectification to control the polarity of thermal diode.

\textit{Magnetic field effect}. In our model, the magnetic field affect the energy gap of the quantum dot.
\begin{figure}[!htbp]
\begin{center}
\includegraphics[width=0.45\textwidth]{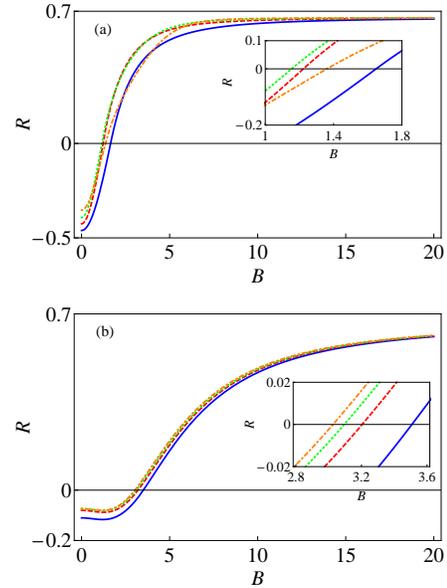}
\end{center}
\caption{\label{fig3}(color online): The relation between thermal rectification and the magnetic field strength in the four different anisotropic situations. Blue solid, red dashed, green dotted and orange dot-dashed correspond to $\chi=0.9$, $\chi=0.3$, $\chi=0.0$ and $\chi=-0.3$ respectively. By setting b=B in Eq.(1), the energy gap of the right dot is zero. The system parameters: the mean temperature $TM=1$, the temperature bias $\Delta T=1$, J=1, $\Gamma_{L}=1$ and $\Gamma_{R}=0.25$. (a) The DM interaction strength is 0. (b) The DM interaction strength is 4. }
\end{figure}
We choose the parameters that the energy gap of the left dot is not zero, and the energy gap of the right one is zero. Different anisotropic system conditions are shown in Fig. \ref{fig3}. The two different situations are described in Fig. \ref{fig3}. (a) and Fig. \ref{fig3}. (b) with the DM interaction strength being zero and four respectively. From the figure, we find that the sign of the thermal rectification changes around certain value of external magnetic field. In different anisotropic situations, the sign of thermal rectification will change at different critical value of external magnetic field. So, we can realize the reversal of the thermal diode model polarity by adjusting the external magnetic field in our two-quantum-dot system, also.

\section{Quantum correlation}
\textit{Quantum entanglement}. To qualify the quantum properties in the steady state, we use the concurrence to measure the quantum entanglement.
\begin{figure}[!htbp]
\begin{center}
\includegraphics[width=0.45\textwidth]{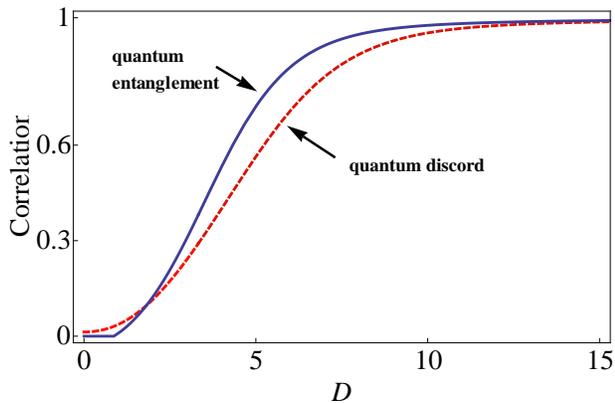}
\end{center}
\caption{\label{fig4}(color online): Quantum correlations with DM interaction. Solid line: quantum entanglement; dashed line: quantum discord. Under the same parameter conditions, quantum entanglement rises with DM interaction more rapidly than quantum discord does. $\Delta T=0.4$, $TM=1$, $\chi=0.3$, B=4, b=2, J=1, $\Gamma_{L}=1$, and $\Gamma_{R}=0.25$.}
\end{figure}
Given the final X states, we have the analytic result of the concurrence $C(\rho^{AB})$ as\cite{W. K. Wooters},
\begin{equation}
\begin{split}
C(\rho^{AB})=2\max\{0,&\frac{J}{2\xi}\sqrt{1+D^{2}}|\rho^{s}_{11}-\rho^{s}_{22}|\\&-\frac{1}{2}\sqrt{(\rho_{33}^{s}+\rho_{44}^{s})^{2}-\frac{B^{2}}{\eta^{2}}(\rho_{33}^{s}
-\rho_{44}^{s})^{2}},\\
&\frac{J\chi}{2\eta}|\rho^{s}_{33}-\rho^{s}_{44}|\\&-\frac{1}{2}\sqrt{(\rho_{11}^{s}+\rho_{22}^{s})^{2}-\frac{b^{2}}{\xi^{2}}(\rho_{11}^{s}-\rho_{22}^{s})^{2}}\}
\end{split}
\end{equation}

\textit{Quantum discord}. Correlation in the quantum aspects of the bipartite system can also be measured by quantum discord. The system quantum discord is defined as\cite{H. Ollivier},
\begin{equation}
D(\rho^{AB})=I(A:B)-C_{cor}(\rho)
\end{equation}

The quantum mutual information $I(A:B)$ has the form as, $I(A:B)=S(\rho_{A})+S(\rho_{B})-S(\rho_{AB})$. The expression $S(\rho)$ is the Von Neumann entropy of the destiny matrix $\rho$, satisfy $S(\rho)=-Tr(\rho\log\rho)$. $\rho^{A}$ and $\rho^{B}$ are the reduced destiny matrix of the system X state density matrix. The Classical correlation $C_{cor}(\rho)$ satisfies the equation $C_{cor}(\rho)=S(\rho^{A})-\min_{B_{i}}[S(\rho|\{B_{i}\})]$, and
the last term $\min_{B_{i}}[S(\rho|\{B_{i}\})]$ is the condition entropy.

\begin{figure}[!htbp]
\begin{center}
\includegraphics[width=0.4\textwidth]{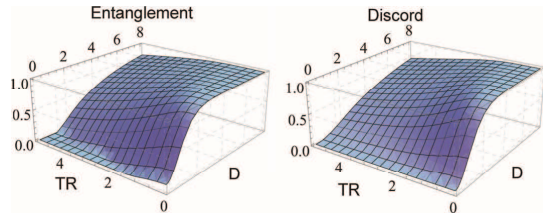}
\end{center}
\caption{\label{fig5} (color online): Quantum correlation with temperature difference of two reservoirs. TR: temperature of right reservoir, D: DM interaction strength. Almost perfect quantum correlation can be achieved even though the temperature difference of two reservoirs is large. Parameters set in calculation: $T_{L}=0.5$, $\chi=0.9$, b=0, B=0.2, J=1, $\Gamma_{L}=1$ and $\Gamma_{R}=0.25$.}
\end{figure}

\begin{figure}[htbp]
\begin{center}
\includegraphics[width=0.4\textwidth]{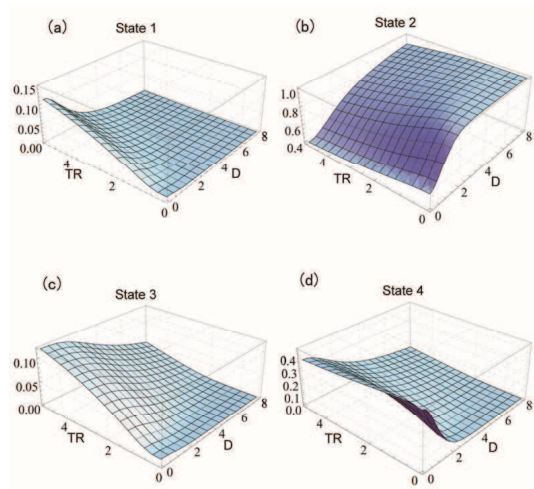}
\end{center}
\caption{\label{fig6} (color online): Population of different system eigenstates $|\varepsilon_{1}\rangle$(a), $|\varepsilon_{2}\rangle$(b), $|\varepsilon_{3}\rangle(c)$ and $|\varepsilon_{4}\rangle$(d) with different right reservoir temperatures and DM interaction strengths. The vertical axis labels the population. Parameters set: $T_{L}=0.5$, $\chi=0.9$, b=0, B=0.2, J=1, $\Gamma_{L}=1$ and $\Gamma_{R}=0.25$.}
\end{figure}
We study the relation between the quantum correlation and the DM interaction strength. From Fig. \ref{fig4}, it is seen that quantum entanglement is more sensitive to the DM interaction strength than quantum discord, and perfect quantum correlation can be got within the small increment of the DM interaction strength.

Almost perfect quantum entanglement and quantum discord can be obtained in the large temperature bias region in our model, see in Fig. \ref{fig5}. The corresponding four system states population are shown in the Fig. \ref{fig6}.The reason for the high value of the quantum correlations is that almost all system population are at the system state $|\varepsilon_{2}\rangle$.

\section{Concluding Remarks}
We study the nonequilibrium two-quantum-dot system with the DM interaction extensively. The behavior of the heat current, quantum entanglement and quantum discord have been investigated. We find that we can control the thermal rectification by two types of modulations, one is adjusting the system DM interaction strength, and the other is regulating the energy gap of two-quantum-dot system. This provides a possible new way to design a polarity-controllable thermal diode. Compared with the spin-chains model studied before\cite{Y. Yan, L. Zhang}, this two-quantum-dot model can offer large thermal rectification when the size is small (N=2). This two-quantum-dot system can be used to build small size phonon devices, and has a potential application in the phonon information.

\section{acknowledgments}
This work was supported in part by
the 10000-Plan of Shandong province, and the National High-Tech Program of China grant No. 2011AA010800 and 2011AA010803, NSFC grant No. 11174177 and 60725416.

{ }

\begin{thebibliography}{ }

\bibitem{A. Nitzan} D. Segal and A. Nitzan, Phys. Rev. Lett. \textbf{94}, 034301 (2005).

\bibitem{B. Li} B. Li, L. Wang, and G. Casati, Phys. Rev. Lett. \textbf{93}, 184301 (2004).

\bibitem{N. B. Li} N. Li, J. Ren, L. Wang, G. Zhang, P. H\"{a}nggi, and B. Li, Rev. Mod. Phys. \textbf{84}, 1045 (2012).

\bibitem{L. Wang} L. Wang and B. Li, Phys. Rev. Lett. \textbf{99}, 177208 (2007).

\bibitem{M. Esposito} M. Esposito, U. Harbola, and S. Mukamel, Rev. Mod. Phys. \textbf{81}, 1665 (2009).

\bibitem{M. Campisi} M. Campisi, P. H\"{a}nggi, and P. Talkner, Rev. Mod. Phys. \textbf{83}, 771 (2011).

\bibitem{Y. Yan} Y. Yan, C. Q. Wu, and B. Li, Phys. Rev. B \textbf{79}, 014207 (2009).

\bibitem{L. Zhang} L. Zhang, Y. Yan, C. Q. Wu, J. S. Wang, and B. Li, Phys. Rev. B \textbf{80}, 172301 (2009).

\bibitem{R. Hanson} R. Hanson and D. D. Awschalom, Nature (London) \textbf{453}, 1043 (2008).

\bibitem{F. Kheirandisha} F. Kheirandisha, S. J. Akhtarshenasb, and H. Mohammadic, Eur. Phys. J. D. \textbf{57}, 129 (2010).

\bibitem{H. Mohammadia} H. Mohammadia, S. J. Akhtarshenas, and F. Kheirandish Eur. Phys. J. D. \textbf{62}, 439 (2011).

\bibitem{L. A. Wu_PRA03} L. A. Wu, and D. A. Lidar, Phys. Rev. A \textbf{67}, 050303(R) (2003).

\bibitem{S. J. Akhtarshenas} F. Kheirandish, S. J. Akhtarshenas, and H. Mohammadi, Phys. Rev. A \textbf{77}, 042309 (2008).

\bibitem{T. Mo} T. Moriya, Phys. Rev. Lett. \textbf{4}, 228 (1960).

\bibitem{T. Moriya} T. Moriya, Phys. Rev. \textbf{120}, 91 (1960).

\bibitem{S. Chutia} S. Chutia, M. Friesen, and R. Joynt, Phys. Rev. B \textbf{73}, 241304(R) (2006).

\bibitem{A. Pfund} A. Pfund, I. Shorubalko, K. Ensslin, and R. Leturcq, Phys. Rev. B \textbf{76}, 161308(R) (2007).

\bibitem{S. Herzog} S. Herzog and M. R. Wegewijs, Nano. Tech. \textbf{21}, 274010 (2010).

\bibitem{L. A. Wu} L. A. Wu, C. X. Yu, and D. Segal, Phys. Rev. E \textbf{80}, 041103 (2009).

\bibitem{L. A} L. A. Wu and D. Segal, Phys. Rev. A \textbf{84}, 012319 (2011).

\bibitem{W. K. Wooters} W. K. Wootters, Phys. Rev. Lett. \textbf{80}, 2245 (1998).

\bibitem{H. Ollivier} H. Ollivier and W. H. Zurek, Phys. Rev. Lett. \textbf{88}, 017901 (2002).

\end{thebibliography}
\end{document}